\documentclass[11pt]{article}
\usepackage[final]{acl}

\usepackage{times}
\usepackage{latexsym}
\usepackage[T1]{fontenc}
\usepackage[utf8]{inputenc}
\usepackage{microtype}
\usepackage{inconsolata}
\usepackage{graphicx}

\usepackage{amsmath}
\usepackage{amssymb}
\usepackage{booktabs}
\usepackage{multirow}
\usepackage{tikz}
\usetikzlibrary{positioning,arrows.meta,calc}
\usepackage{algorithm}
\usepackage{algpseudocode}

\title{DistilVDR: A Compact End-to-End Visual Document Retriever\\
  via Dual-Student Distillation}

\author{
  Zhuchenyang Liu\textsuperscript{1} \quad
  Ziyi Wang\textsuperscript{2} \quad
  Yao Zhang\textsuperscript{1} \quad
  Yu Xiao\textsuperscript{1} \\
  \textsuperscript{1}Aalto University, Finland \\
  \textsuperscript{2}Independent Researcher, Netherlands \\
  \texttt{zhuchenyang.liu@aalto.fi}
}

\begin{document}
\maketitle

\begin{abstract}
Visual document retrieval (VDR) is dominated by multi-billion-parameter models that are slow to index at full corpus scale and expensive to serve. Prior compression routes either train a smaller multi-vector encoder from scratch or distil only the query side; neither yields a compact single-vector retriever end-to-end. We present DistilVDR, a 524\,M end-to-end VDR system distilled bilaterally from a single 8\,B vision-language teacher under a pointwise cosine alignment loss. All supervision comes from the frozen teacher's embedding space, which was itself trained with relevance supervision, so the student objective needs no relevance labels, negative sampling, or contrastive term. We match VDR's text-query and image-document input asymmetry with an asymmetric encoder-only student that concentrates visual capacity on the document side and keeps the query side at 70\,M parameters. We release two variants that share the same encoders and training and differ only in the document encoder's visual-tile budget: DistilVDR-HiRes attains 61.74 average NDCG@5 on ViDoRe v1+v2+v3 (86.9\,\% of the 8\,B teacher) and leads every reproduced sub-1\,B baseline on the high-resolution-sensitive v3 benchmark, while DistilVDR-Fast attains 59.98 at a 3$\times$ smaller visual-token budget. Both variants store one million documents in a 15.6 times smaller index than the strongest sub-1\,B multi-vector baseline and index the corpus an order of magnitude faster. The code is available at \url{https://github.com/Ryenhails/NanoVDR}.
\end{abstract}

\section{Introduction}
\label{sec:intro}

Enterprise document search, technical documentation lookup, and
visual retrieval-augmented generation \cite{yu2025visrag} all
require retrieving relevant document pages from a corpus given a
text query.  Visual document retrieval (VDR) addresses this task.
VDR encodes each page directly as an image.  This avoids the errors
of an Optical Character Recognition (OCR) or layout extraction
pipeline and preserves tables, figures, and layout structure.

State-of-the-art VDR systems are heavy.  Top-performing retrievers
span two to eight billion parameters across both single-vector
and multi-vector paradigms; examples include the single-vector
Qwen3-VL-Embedding-8B \cite{li2026qwen3} and the multi-vector
Tomoro-ColQwen3-8B \cite{tomoro2025colqwen3}.  These models
require more than 16~GB of GPU memory to embed a single document.
Indexing a million-document corpus with such a model costs tens
of GPU-hours.

Two lines of work have addressed this cost.  The first line trains a
smaller encoder from scratch and combines it with multi-vector late
interaction
\cite{khattab2020colbert,santhanam-etal-2022-colbertv2,santhanam2022plaid}.
Per-token storage compensates for the reduced parameter count.
Representative models include colSmol-500M \cite{vidore2024colsmol},
SauerkrautLM-ColLFM2-450M \cite{vago2025collfm2}, and ModernVBERT
\cite{teiletche2025modernvbert}, all built on the ColPali
architecture \cite{faysse2025colpali}.  Among these, ModernVBERT
departs from the ColPali decoder lineage with an encoder-only design.
It combines a SigLIP2 visual encoder
\cite{tschannen2025siglip2multilingualvisionlanguage} with a
ModernBERT text backbone \cite{warner2025smarter}.
\citet{teiletche2025modernvbert} report that this encoder-only
design outperforms an equivalent causal decoder on document
retrieval by 10.6 NDCG@5.

However, all systems in this line pay the multi-vector cost.
Per-token storage inflates the index by an order of magnitude, and
late-interaction MaxSim scoring inflates per-query latency by two
orders of magnitude relative to a single dot product.  A body of work
attacks these costs post hoc, by pruning or merging patches in an
already-built index
\cite{ma2025storageefficientvisualdocumentretrieval,yan2026sculptingvectorspaceefficient},
by fusing internal transformer layers into a more discriminative
embedding \cite{li2026minerminingmultimodalinternal}, or by
reranking single-vector candidates with a multi-vector model
\cite{kim2026hybridvectorretrievalvisuallyrich}.  All of it
presumes an already-trained strong base model, so the training and
deployment cost of the underlying multi-billion-parameter
retriever remains.  Going single-vector within the same
architecture removes the index cost but loses quality:
BiModernVBERT is 17.6 NDCG@5 below ColModernVBERT on ViDoRe v1 and
20.3 below it on v2 \cite{teiletche2025modernvbert}.  A small
encoder trained from scratch with contrastive learning therefore
struggles to produce a single-vector representation that matches
multi-vector retrieval quality.

The second line distils a small student from a large teacher.
This is a well-established paradigm in text retrieval.  A heavier
teacher, typically a cross-encoder reranker or a larger bi-encoder
\cite{hofstatter2021tasb,chen-etal-2024-m3}, supervises a lighter
bi-encoder student, which inherits much of the teacher's quality at a
fraction of the inference cost.  In visual document retrieval, the
only prior work in this direction, to our knowledge, is NanoVDR
\cite{liu2026nanovdr}.  A 70\,M DistilBERT
\cite{sanh2019distilbert} query encoder is trained to match the
embedding space of a 2\,B vision-language teacher with a cosine
alignment loss.  The student matches the teacher within a few NDCG
points despite being thirty times smaller, which shows that
distillation is a viable route to compact single-vector VDR.

Neither line produces a compact single-vector retriever end-to-end.
The from-scratch multi-vector line yields a small encoder but pays an
order of magnitude in index size and two orders of magnitude in
scoring latency.  The distillation line yields a small query encoder
but leaves the document encoder at the size of the teacher.  NanoVDR
keeps the 2\,B teacher in the document path, so indexing a
million-document corpus still costs tens of GPU-hours and every
deployment must host the teacher in memory.

We close this gap with DistilVDR, a 524\,M end-to-end
single-vector visual document retriever.  Our contribution is a
system: (i) a bilateral cosine alignment distillation setup in
which each student regresses onto cached embeddings of a single
8\,B vision-language teacher~\cite{li2026qwen3}; (ii) an
asymmetric encoder-only student matched to VDR's input asymmetry,
pairing a heavy visual-and-text document tower with a lightweight
70\,M text-only query tower; and (iii) a unified reproduction of
twelve released retrievers from 250\,M to 8.8\,B under one
evaluation and one profiling pipeline, reporting retrieval quality
together with query latency, indexing throughput, peak VRAM, index
footprint, and scoring latency.  Composed under one teacher, these
yield a retriever that dominates the from-scratch small
multi-vector route at the 500\,M scale on quality and on every
deployment axis at once.
The two deployment variants we release differ only in the document
encoder's visual-tile budget.  DistilVDR-HiRes reaches 61.74
average NDCG@5 on ViDoRe v1+v2+v3
\cite{faysse2025colpali,mace2025vidore,loison2026vidore}, 86.9\,\%
of the 8\,B teacher, and leads every reproduced sub-1\,B baseline
on the high-resolution-sensitive v3 benchmark.  At a $3\times$
smaller visual-token budget, DistilVDR-Fast still exceeds every
reproduced sub-1\,B baseline, at 59.98 average NDCG@5.
Both variants store one million documents in a 15.6 times
smaller index than the strongest sub-1\,B multi-vector baseline
and index the corpus an order of magnitude faster.

\section{Related Work}
\label{sec:related}

Section~\ref{sec:intro} covers the directly competing visual
document retrievers; here we situate our work against three
adjacent areas.

\paragraph{Distillation in dense text retrieval.}
Hard-negative mining
\cite{xiong2020approximatenearestneighbornegative,qu-etal-2021-rocketqa},
retrieval-oriented pretraining \cite{xiao-etal-2022-retromae}, and
cross-encoder-to-bi-encoder distillation
\citep{hofstatter2021tasb,chen-etal-2024-m3} are common building
blocks of modern dense retrievers
\cite{wang2024textembeddingsweaklysupervisedcontrastive,wang2024multilinguale5textembeddings,sturua2024jinaembeddingsv3multilingualembeddingstask}.
The asymmetric pattern in which a heavy reranker teaches a
lightweight bi-encoder is the closest text-side analogue to our
setup, and differs from it in two ways.  Our teacher produces a
single image embedding rather than a query-document relevance
score, so the distillation target is geometric rather than a
scalar; and our asymmetry is across modalities rather than across
model classes, with a text-only query side and an image document
side.

\paragraph{Universal multimodal embedders.}
A second line fine-tunes a large vision-language model into a
universal embedder over arbitrary modality pairs
\cite{jiang2024vlm2vec,wei2024uniir}.  We share the idea of
converting a generative vision-language model into an embedding
model, but trade their breadth at multi-billion scale for depth in
one subdomain at sub-1\,B scale.  The fixed input shape of VDR,
text queries against image documents, lets both encoders stay
small.

\paragraph{OCR-free document understanding.}
A third line feeds document images directly to the model.  Donut
\cite{10.1007/978-3-031-19815-1_29}, Pix2Struct
\cite{pmlr-v202-lee23g} and UDOP \citep{Tang_2023_CVPR} decode text
outputs; LayoutLM and LayoutLMv3
\citep{xu2020layoutlm,huang2022layoutlmv3} are bidirectional
encoders with explicit layout tokens.  They share our premise that
direct image input avoids OCR error compounding, but target
generation or extraction and produce representations not designed
for nearest-neighbour retrieval.  We produce one dense vector per
page that indexes with standard dense-retrieval infrastructure.

\section{Method}
\label{sec:method}

\subsection{Problem setup}
\label{sec:method:setup}

Given a text query $q$ and a corpus of document images
$\{d_1, \ldots, d_N\}$, the system ranks the corpus by relevance to
$q$.  We work in the single-vector dense retrieval setting throughout.
A query encoder $f_q$ maps the query to a vector
$\mathbf{q} \in \mathbb{R}^k$ and a document encoder $f_d$ maps each
document image to a vector $\mathbf{d}_i \in \mathbb{R}^k$ in the
same space.  Both vectors are L2-normalised.  The retrieval score is
the dot product $s(q, d_i) = \mathbf{q}^\top \mathbf{d}_i$.  Document
vectors are computed once at indexing time and queried with any
standard dense-retrieval engine.

\subsection{Dual-student distillation}
\label{sec:method:arch}

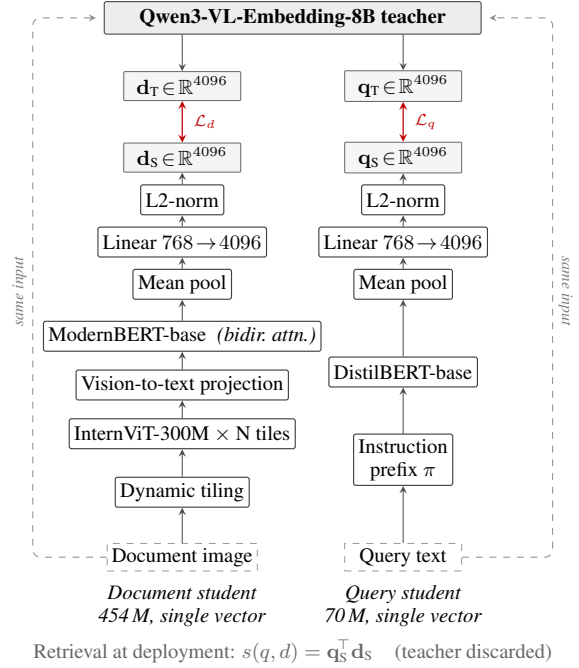
\begin{figure}[!t]
  \centering
  \resizebox{\columnwidth}{!}{

\begin{tikzpicture}[
  font=\small,
  >=stealth,
  node distance=4mm,
  block/.style={
    rectangle, draw=black!75, line width=0.5pt,
    rounded corners=1pt, fill=white,
    inner sep=2.2pt, minimum height=4.6mm, align=center,
  },
  teacherbox/.style={
    rectangle, draw=black!70, line width=0.5pt,
    rounded corners=1pt, fill=black!7,
    inner sep=3pt, minimum height=5mm, minimum width=58mm, align=center,
  },
  emb/.style={
    rectangle, draw=black!60, fill=black!4,
    rounded corners=0.5pt, inner sep=2pt,
    minimum height=4.3mm, minimum width=18mm, align=center,
  },
  io/.style={
    rectangle, draw=black!40, fill=white, dashed,
    inner sep=2pt, minimum height=4.2mm, minimum width=18mm, align=center,
  },
  arrow/.style={->, line width=0.45pt, draw=black!70},
  loss/.style={<->, line width=0.55pt, draw=red!75!black},
  teacherin/.style={->, line width=0.4pt, draw=black!45, dashed},
]

\node[teacherbox] (teacher)
  at (0, 2.6) {\textbf{Qwen3-VL-Embedding-8B teacher}};

\node[emb] (dt) at (-1.7, 1.55) {$\mathbf{d}_{\text{T}}\!\in\!\mathbb{R}^{4096}$};
\node[emb] (qt) at ( 1.7, 1.55) {$\mathbf{q}_{\text{T}}\!\in\!\mathbb{R}^{4096}$};
\draw[arrow] (teacher.south -| dt.north) -- (dt.north);
\draw[arrow] (teacher.south -| qt.north) -- (qt.north);

\node[emb] (ds) at (-1.7, 0.45) {$\mathbf{d}_{\text{S}}\!\in\!\mathbb{R}^{4096}$};
\node[emb] (qs) at ( 1.7, 0.45) {$\mathbf{q}_{\text{S}}\!\in\!\mathbb{R}^{4096}$};

\draw[loss] (ds) -- node[right=1.2pt, font=\scriptsize, text=red!75!black]
                    {$\mathcal{L}_d$} (dt);
\draw[loss] (qs) -- node[right=1.2pt, font=\scriptsize, text=red!75!black]
                    {$\mathcal{L}_q$} (qt);

\node[block] (norm_d) at (-1.7, -0.20) {L2-norm};
\node[block] (proj_d) at (-1.7, -0.85) {Linear $768\!\to\!4096$};
\node[block] (pool_d) at (-1.7, -1.50) {Mean pool};
\node[block] (modern)  at (-1.7, -2.30) {ModernBERT-base \;\textit{(bidir.\ attn.)}};
\node[block] (connect) at (-1.7, -3.05) {Vision-to-text projection};
\node[block] (intern)  at (-1.7, -3.80) {InternViT-300M $\times$ N tiles};
\node[block] (tile) at (-1.7, -4.70) {Dynamic tiling};
\node[io] (doc_img) at (-1.7, -5.70) {Document image};

\draw[arrow] (doc_img) -- (tile);
\draw[arrow] (tile)    -- (intern);
\draw[arrow] (intern)  -- (connect);
\draw[arrow] (connect) -- (modern);
\draw[arrow] (modern)  -- (pool_d);
\draw[arrow] (pool_d)  -- (proj_d);
\draw[arrow] (proj_d)  -- (norm_d);
\draw[arrow] (norm_d)  -- (ds);

\node[block] (norm_q) at ( 1.7, -0.20) {L2-norm};
\node[block] (proj_q) at ( 1.7, -0.85) {Linear $768\!\to\!4096$};
\node[block] (pool_q) at ( 1.7, -1.50) {Mean pool};
\node[block] (distil) at ( 1.7, -2.85) {DistilBERT-base};
\node[block, align=center]
       (qprefix)      at ( 1.7, -4.20) {Instruction\\prefix $\pi$};
\node[io] (q_text)    at ( 1.7, -5.70) {Query text};

\draw[arrow] (q_text)  -- (qprefix);
\draw[arrow] (qprefix) -- (distil);
\draw[arrow] (distil)  -- (pool_q);
\draw[arrow] (pool_q)  -- (proj_q);
\draw[arrow] (proj_q)  -- (norm_q);
\draw[arrow] (norm_q)  -- (qs);

\node[align=center, font=\footnotesize] at (-1.7, -6.45)
  {\textit{Document student}\\\textit{454\,M, single vector}};
\node[align=center, font=\footnotesize] at ( 1.7, -6.45)
  {\textit{Query student}\\\textit{70\,M, single vector}};

\draw[teacherin, rounded corners=1.2mm, shorten >=1.2mm]
  (doc_img.west) -- (-4.00, -5.70) -- (-4.00, 2.6) -- (teacher.west);
\draw[teacherin, rounded corners=1.2mm, shorten >=1.2mm]
  (q_text.east) -- ( 4.00, -5.70) -- ( 4.00, 2.6) -- (teacher.east);
\node[font=\scriptsize, text=black!50, rotate=90]
    at (-4.20, -1.6) {\textit{same input}};
\node[font=\scriptsize, text=black!50, rotate=-90]
    at ( 4.20, -1.6) {\textit{same input}};

\node[align=center, font=\footnotesize, text=black!60]
  at (0, -7.15) {Retrieval at deployment: $s(q,d)=\mathbf{q}_{\text{S}}^{\!\top}\mathbf{d}_{\text{S}}$
                 \;\; (teacher discarded)};

\end{tikzpicture}}
  \caption{Dual-student distillation architecture.  The teacher
    encodes the image (left) and the query (right); each student
    is trained independently to match its teacher target.}
  \label{fig:arch}
\end{figure}

The architecture is a two-student instance of the asymmetric
distillation paradigm.  A frozen 8\,B vision-language teacher
\cite{li2026qwen3} produces target embeddings of dimension $k = 4096$
for both modalities.  Two students learn to reproduce these targets
independently.  The query student $f_q$ takes a text query and the
document student $f_d$ takes a document image.  Both students project
to the teacher's output space and are L2-normalised.  Retrieval at
deployment uses the students only; the teacher is discarded after
training.  Figure~\ref{fig:arch} illustrates the architecture and
the two distillation objectives.

\subsection{Document encoder}
\label{sec:method:doc}

The document encoder $f_d$ has 454\,M parameters
(300\,M InternViT visual encoder + 150\,M ModernBERT-base text
backbone + 4\,M visual-to-text projection and final 768$\to$4096
output head) and proceeds in four stages.

First, DistilVDR enforces a fixed per-document visual-token
budget.  We cap the per-page tile count at $T_{\max}$, a
release-time design choice that distinguishes the Fast and
HiRes variants (Section~\ref{sec:exp:setup}), and append a
single thumbnail at the visual encoder's native resolution for
global context.  Given $T_{\max}$ and the page's aspect ratio,
we choose an aspect-ratio-matched grid layout at the same
native tile resolution, following the InternVL-V2 partition
rule~\cite{chen2025expandingperformanceboundariesopensource};
Algorithm~\ref{alg:tile} in Appendix~\ref{app:tiling}
formalises the procedure.  The cap resolves the two failure modes
of document encoding at once: a single low-resolution view loses
small text, table cells, and figure annotations, while a single
high-resolution view exceeds the text backbone's context window.

Second, every tile is encoded by an InternViT-300M-448 visual encoder
\cite{chen2024internvl} into 1\,024 patch tokens of dimension 768.
We select InternViT-300M for two reasons: its 448-pixel native tile
resolution matches our dynamic-tiling rule without any patch
resampling, and its document-rich vision-language pre-training
transfers well to page-level document imagery.
Patch tokens from all tiles are concatenated into one sequence of
length up to 7\,168.

Third, the visual sequence is mapped into the embedding space of a
ModernBERT-base text backbone \cite{warner2025smarter} by a learned
linear projection.  ModernBERT then re-encodes the projected
visual tokens with bidirectional attention, with no text tokens in
the input.  We use the text backbone here as a contextual encoder
over visual tokens.  This usage follows ModernVBERT
\cite{teiletche2025modernvbert}, which shows that an encoder with
bidirectional attention outperforms a causal decoder on document
retrieval.  We select ModernBERT-base because its 8\,192-token
context window absorbs the worst-case 7\,168-token sequence from
seven tiles without truncation, and because its rotary positional
embeddings and alternating local/global attention yield strong
long-context representations at the 150\,M parameter scale.

Fourth, the contextualised tokens are mean pooled, projected from 768
to 4096 dimensions by a final linear layer, and L2-normalised.

\subsection{Query encoder}
\label{sec:method:query}

The query encoder $f_q$ has 70\,M parameters and proceeds in three
stages.  First, the query text is prefixed by the same instruction
string $\pi$ that the teacher was trained with, so that the student
input matches the teacher input at distillation time.  Second, the
prefixed sequence is encoded by a DistilBERT-base text encoder
\cite{sanh2019distilbert} and mean pooled over the contextual token
representations.  Third, the pooled vector is
passed through a linear projection from 768 to 4096 dimensions and
L2-normalised.

\subsection{Distillation objective}
\label{sec:method:loss}

Let $T$ denote the frozen teacher.  For a document image $d$, the
teacher produces a target vector $T(d) \in \mathbb{R}^{4096}$.  For a
text query $q$ with instruction prefix $\pi$, it produces a target
$T(\pi \!\circ\! q) \in \mathbb{R}^{4096}$.  Both targets are
L2-normalised by construction.  Each student is trained independently
against these cached targets under a cosine alignment loss,
\begin{align}
  \mathcal{L}_d &= 1 - \langle f_d(d),\; T(d) \rangle, \label{eq:ld}\\
  \mathcal{L}_q &= 1 - \langle f_q(\pi \!\circ\! q),\;
                              T(\pi \!\circ\! q) \rangle, \label{eq:lq}
\end{align}
where $\langle \cdot,\cdot \rangle$ denotes the dot product between
L2-normalised vectors.  The two students never share a forward
pass during training, so the doc-side and query-side distillations
are fully decoupled.

No contrastive term, hard negative, or relevance label enters the
student objective.  The supervision is not label-free in an
absolute sense, since the teacher's embedding space was itself
trained with relevance supervision.  The objective removes
relevance signals and negative sampling \emph{at student training
time}, which is what keeps the two distillations decoupled and
independently parallelisable.  Implementation details,
datasets, and hyperparameters are deferred to
Section~\ref{sec:exp}.

\section{Experiments}
\label{sec:exp}

\subsection{Experimental setup}
\label{sec:exp:setup}

\paragraph{Benchmarks and metric.}
We evaluate on the full 22-dataset ViDoRe suite: v1
\cite{faysse2025colpali} (10 datasets, English), v2
\cite{mace2025vidore} (4, multilingual) and v3
\cite{loison2026vidore} (8, professional domains), each dataset
listed in Appendix~\ref{app:vidore}.  We report NDCG@5, computed
with \texttt{pytrec\_eval} and averaged within each benchmark.

\paragraph{Training data and recipe.}
The document mixture is 1.20\,M unique images from public,
permissively licensed sources, assembled in three parts: a 711\,K
base mixture, a 454\,K multi-domain supplement, and a 31.7\,K
finance supplement, spanning scientific publications, multilingual
web PDFs, industrial and regulatory reports, and financial
filings.  All three are deduplicated by perceptual hash against
one another and against all three ViDoRe evaluation corpora
(Appendix~\ref{app:data}).  The query mixture is the
1.49\,M-query NanoVDR training set \cite{liu2026nanovdr}.
Both students are trained with AdamW under a one-cycle schedule
against cached 4096-d teacher targets.  Appendices~\ref{app:data}
and~\ref{app:recipe} give the per-source breakdown and the full
hyperparameters.  The document encoder is trained in two variants
differing only in the tile cap of Section~\ref{sec:method:doc}:
DistilVDR-HiRes (six 448$\times$448 tiles plus thumbnail;
7\,168-token worst case) and DistilVDR-Fast (two tiles plus
thumbnail; 3\,072-token worst case).  Data, optimizer, and epoch
budget are identical across the two variants and the query
encoder.

\paragraph{Teacher precomputation cost.}
The teacher runs exactly once, before any student training.
Caching targets for the 1.20\,M images and 1.49\,M queries cost
99.5 H200-GPU-hours across three sharded batches, one per part of
the mixture: 40.5 for the 711\,K base mixture together with its
queries, their translations, and the validation split; 58.1 for
the multi-domain supplement, as encoded before deduplication
reduced it to 454\,K images; and 0.9 for the finance supplement,
at a measured 11.3 images/sec.  Every result in this
paper, including all retrained ablations, reuses these cached
targets, and this is the only point at which the 8\,B model is
executed.

\paragraph{Baselines and protocol.}
We reproduce twelve publicly released retrievers spanning
single-vector and multi-vector designs from 250\,M to 8.8\,B
parameters, plus the 8\,B teacher as a single-vector oracle;
Appendix~\ref{app:baselines} names each one.  Every system is
loaded from its official release and run inside one evaluation
driver, so input formatting, deduplication, and metric computation
are identical across systems.  Multi-vector baselines use their
authors' recommended late-interaction routine; single-vector
baselines use brute-force dot product over the full corpus.

\begin{table*}[t]
  \centering
  \caption{Retrieval quality (NDCG@5) on ViDoRe v1, v2 and v3.
    ``Type'' is single vector or multi-vector with MaxSim.}
  \label{tab:main}
  \small
  \begin{tabular}{@{}lrlcccc@{}}
    \toprule
    Model & Params & Type & v1 & v2 & v3 & Avg \\
    \midrule
    \multicolumn{7}{@{}l}{\textit{Sub-1\,B}} \\
    SigLIP2-L                                       &  880\,M & single & 43.58 & 20.17 & 14.04 & 25.93 \\
    BiModernVBERT                                   &  250\,M & single & 37.40 & 10.88 &  5.52 & 17.93 \\
    colSmol-256M                                    &  256\,M & multi  & 79.72 & 34.63 & 25.23 & 46.53 \\
    colSmol-500M                                    &  478\,M & multi  & 82.42 & 43.09 & 33.52 & 53.01 \\
    ColModernVBERT                                  &  250\,M & multi  & 76.76 & 33.18 & 17.45 & 42.46 \\
    SauerkrautLM-ColLFM2                            &  451\,M & multi  & 78.24 & 45.09 & 33.19 & 52.17 \\
    \textbf{DistilVDR-Fast (ours)}  & \textbf{524\,M} & \textbf{single}
                                                 & 81.34
                                                 & 54.95
                                                 & 43.66
                                                 & 59.98 \\
    \textbf{DistilVDR-HiRes (ours)}  & \textbf{524\,M} & \textbf{single}
                                                 & \textbf{82.81}
                                                 & \textbf{55.34}
                                                 & \textbf{47.07}
                                                 & \textbf{61.74} \\
    \midrule
    \multicolumn{7}{@{}l}{\textit{Mid- to large-scale references}} \\
    DSE-Qwen2                 & 2.2\,B & single & 85.14 & 55.70 & 41.28 & 60.71 \\
    Qwen3-VL-Embedding-2B     & 2.1\,B & single & 84.30 & 65.25 & 49.98 & 66.51 \\
    ColPali v1.3              & 2.9\,B & multi  & 84.21 & 54.72 & 42.04 & 60.32 \\
    Tomoro-ColQwen3-4B        & 4.4\,B & multi  & 90.22 & 65.25 & 57.57 & 71.01 \\
    ColNomic-7B               & 7.0\,B & multi  & 89.76 & 60.44 & 55.87 & 68.69 \\
    Tomoro-ColQwen3-8B        & 8.8\,B & multi  & 90.61 & 65.00 & 59.00 & 71.54 \\
    Qwen3-VL-Embedding-8B (teacher)
                              & 8.1\,B & single & 87.31 & 69.76 & 56.07 & 71.05 \\
    \bottomrule
  \end{tabular}
\end{table*}

\subsection{Retrieval quality}
\label{sec:exp:quality}

Table~\ref{tab:main} reports NDCG@5 on the full ViDoRe suite for
every reproduced baseline alongside the two DistilVDR variants.
Against the strongest reproduced sub-1\,B baseline, colSmol-500M
at 53.01 average, HiRes leads by 8.73 points and Fast by 6.97.
Both variants exceed every other sub-1\,B baseline by more.

The two variants separate on the hardest benchmark.  On v3, built
on long professional reports with dense small text and complex
layouts, HiRes scores 47.07 against Fast's 43.66, while on v1 and
v2 they stay within 1.5 points of each other.  HiRes leads the
next-best sub-1\,B retriever on v3 by 13.55 points (47.07 vs.\
33.52); Fast trades that v3 quality for a $3\times$ smaller
visual-token budget, which Section~\ref{sec:exp:efficiency} turns
into deployment cost.

Against larger references the variants are competitive at 2--3\,B
and behind at 4--8\,B.  HiRes exceeds DSE-Qwen2 (2.2\,B) by 1.03
and ColPali v1.3 (2.9\,B) by 1.42 average points despite being
four to six times smaller, and Fast trails both by under one
point.  Both trail Qwen3-VL-Embedding-2B, and the 4--8\,B
multi-vector models remain 6.95 to 9.80 points ahead of HiRes,
with the gap concentrated on v2 and v3.  The 8\,B teacher exceeds
HiRes by 9.31 points; HiRes and Fast retain 86.9\,\% and 84.4\,\%
of its average NDCG@5.

\paragraph{Gap decomposition.}
To localise the residual gap between DistilVDR-HiRes and the
8\,B teacher we evaluate the four combinations of student and
teacher on each side.
Table~\ref{tab:gap_decomp} reports the resulting NDCG@5.  Replacing
the document side with the student costs 6.03 NDCG@5 on average
(T$\times$T vs.\ T$\times$S).  Replacing the query side with the
student costs 4.69 points on average (T$\times$T vs.\ S$\times$T).
The two side losses do not simply add to the full gap of 9.31 points;
the residual is the interaction between the two student errors.

\begin{table}[!t]
  \centering
  \caption{Gap decomposition (NDCG@5) for DistilVDR-HiRes.
    T = teacher, S = student.  Bottom row is the full system.}
  \label{tab:gap_decomp}
  \small
  \setlength{\tabcolsep}{4pt}
  \begin{tabular}{@{}lrrrr@{}}
    \toprule
    Query $\times$ Doc & v1 & v2 & v3 & Avg \\
    \midrule
    T $\times$ T \;(oracle) & 87.31 & 69.76 & 56.07 & 71.05 \\
    T $\times$ S            & 83.72 & 60.92 & 50.43 & 65.02 \\
    S $\times$ T            & 84.68 & 64.30 & 50.09 & 66.36 \\
    S $\times$ S \;(ours)   & 82.81 & 55.34 & 47.07 & 61.74 \\
    \bottomrule
  \end{tabular}
\end{table}

\paragraph{Where the residual gap lives.}
One 4096-d vector per page already expresses retrieval quality
well beyond what our student reaches.  The teacher is itself a
single-vector retriever, and at 71.05 average NDCG@5 it is level
with the 4.4\,B multi-vector Tomoro-ColQwen3-4B (71.01) and above
the 7\,B ColNomic-7B (68.69).  DistilVDR is therefore limited by
capacity within the single-vector format rather than by the format
itself.  The document side accounts for 6.03 of the 9.31-point gap
(Table~\ref{tab:gap_decomp}), and shrinking the output vector from
4096 to 768 dimensions alone costs 4.77 NDCG@5 on v3
(Table~\ref{tab:abl_combined}, block (c)).

\begin{table*}[t]
  \centering
  \caption{Deployment efficiency on a single H200 GPU at fixed
    batch size $B=8$ in bf16.  Each encoder runs under its
    best-supported flash-attention backend~\cite{dao2022flashattention,dao2024flashattention};
    per-baseline details are in Appendix~\ref{app:effbench}.
    Index size is per million documents.  Avg NDCG@5 from
    Table~\ref{tab:main} is repeated for reference.  In each
    efficiency column we \textbf{bold} the column-best and
    \underline{underline} the second-best.}
  \label{tab:enc}
  \footnotesize
  \setlength{\tabcolsep}{3pt}
  \begin{tabular}{@{}lrlrrrrrr@{}}
    \toprule
    Model & Params & Type
      & \begin{tabular}{@{}r@{}}Query\\(ms)\end{tabular}
      & \begin{tabular}{@{}r@{}}Doc thpt\\(docs/s)\end{tabular}
      & \begin{tabular}{@{}r@{}}Peak VRAM\\(GB)\end{tabular}
      & \begin{tabular}{@{}r@{}}Index\\/ 1\,M\end{tabular}
      & \begin{tabular}{@{}r@{}}Score 10\,K\\(ms)\end{tabular}
      & \begin{tabular}{@{}r@{}}Avg\\NDCG@5\end{tabular} \\
    \midrule
    \multicolumn{9}{@{}l}{\textit{Sub-1\,B}} \\
    SigLIP2-L                                 &  880\,M & single & 113.8           & 28.22           & \textbf{1.98} & \underline{4.1\,GB} & \underline{1.3} & 25.93 \\
    BiModernVBERT                             &  250\,M & single & \underline{7.4} &  2.49           &  7.47         & \textbf{3.1\,GB}    & \textbf{0.9}    & 17.93 \\
    colSmol-256M                              &  256\,M & multi  &  23.6           &  2.58           &  4.49         & 256\,GB             & 1\,259          & 46.53 \\
    colSmol-500M                              &  478\,M & multi  &  22.6           &  2.82           &  4.97         & 256\,GB             & 1\,161          & 53.01 \\
    ColModernVBERT                            &  250\,M & multi  &  61.8           &  3.06           &  4.26         & 256\,GB             & 1\,238          & 42.46 \\
    SauerkrautLM-ColLFM2                      &  451\,M & multi  &   8.5           & 19.02           &  2.56         & 256\,GB             & 1\,330          & 52.17 \\
    \textbf{DistilVDR-Fast (ours)}            &  524\,M & single & \textbf{3.4}    & \textbf{99.04}  & \underline{2.10} & 16.4\,GB         & 9.6             & 59.98 \\
    \textbf{DistilVDR-HiRes (ours)}           &  524\,M & single & \textbf{3.4}    & \underline{36.82} & 3.07        & 16.4\,GB            & 9.6             & 61.74 \\
    \midrule
    \multicolumn{9}{@{}l}{\textit{Mid- to large-scale references}} \\
    DSE-Qwen2                                 & 2.2\,B & single & 167.4 & 17.17 &  6.31 &  6.1\,GB & 2.2     & 60.71 \\
    Qwen3-VL-Embedding-2B                     & 2.1\,B & single &  14.5 &  8.53 &  7.03 &  8.2\,GB & 3.4     & 66.51 \\
    ColPali v1.3                              & 2.9\,B & multi  & 266.7 & 17.24 &  7.81 & 264\,GB  & 1\,158  & 60.32 \\
    Tomoro-ColQwen3-4B                        & 4.4\,B & multi  & 266.4 & 11.91 & 12.93 & 819\,GB  & 3\,187  & 71.01 \\
    ColNomic-7B                               & 7.0\,B & multi  & 542.6 &  9.96 & 17.88 & 256\,GB  & 1\,206  & 68.69 \\
    Tomoro-ColQwen3-8B                        & 8.8\,B & multi  & 499.3 &  9.20 & 21.76 & 819\,GB  & 3\,176  & \textbf{71.54} \\
    Qwen3-VL-Embedding-8B (teacher)           & 8.1\,B & single &  19.8 &  5.40 & 19.28 & 16.4\,GB & 9.4     & \underline{71.05} \\
    \bottomrule
  \end{tabular}
\end{table*}

\subsection{Efficiency analysis}
\label{sec:exp:efficiency}

Table~\ref{tab:enc} reports throughput and resource usage for
the twelve baselines and the distillation teacher of
Table~\ref{tab:main}.  Every encoder runs
at fixed batch size $B=8$ in bf16 under its best-supported
flash-attention backend~\cite{dao2022flashattention,dao2024flashattention}
on a single H200 GPU; per-baseline implementation details are
deferred to Appendix~\ref{app:effbench}.  The four 250--500\,M
baselines (BiModernVBERT, colSmol-256M, colSmol-500M,
ColModernVBERT) all use the uncapped Idefics3 image
splitter~\cite{laurenccon2024building}, which emits more than ten
sub-images per document page; DistilVDR caps its budget at seven
(Section~\ref{sec:method:doc}), so its image tower runs over a
shorter visual sequence at comparable parameter scales.

\paragraph{Encoding and indexing.}
DistilVDR-Fast attains the highest document throughput at 99.04
docs/sec and 2.10\,GB peak VRAM, an order of magnitude faster than
every multi-vector baseline at any scale, several times faster
than every other sub-1\,B single-vector baseline, and roughly
18$\times$ faster than the 8\,B teacher.  HiRes returns part of
that throughput for its larger tile budget and still runs
7$\times$ faster than the teacher.  SigLIP2-L is the only sub-1\,B
baseline with lower peak VRAM than Fast.  Both variants encode
queries in 3.4\,ms through the text-only DistilBERT path, faster
than every other profiled system.

\paragraph{Index storage and scoring.}
The multi-vector index footprint is the per-document token count
times the embedding dimension times the storage precision.  Sub-1\,B
multi-vector baselines store roughly 1\,000 128-d
\texttt{float16} vectors per page (256\,GB per million documents)
and the 4--8\,B Tomoro baselines 1\,280 320-d tokens (819\,GB).
DistilVDR stores one 4096-d \texttt{float32} vector, 16.4\,GB, and
scores 10\,000 documents in 9.6\,ms against the 1.1--3.2\,s the
multi-vector baselines need for the same MaxSim.  An end-to-end
multi-vector deployment at sub-1\,B scale therefore costs roughly
16$\times$ in storage and 100$\times$ in scoring latency relative
to DistilVDR.

\section{Ablations}
\label{sec:ablation}

Table~\ref{tab:abl_combined} collects four ablations: the visual
tile budget, the training-data scale, and the projection-head
output dimension on the document side, plus the query-encoder
backbone.  Every variant is retrained from scratch under the
recipe of Appendix~\ref{app:recipe}, changing one factor at a
time.  Blocks (a) and (b) are scored end to end, student query
against student document.  Blocks (c) and (d) are scored under
single-side isolation, pairing the retrained tower with the
teacher's other tower so that the factor under study is not
confounded by the other tower's error; their absolute values are
therefore higher than those of the end-to-end system of
Section~\ref{sec:exp:quality} and are comparable only within a
block.  A fifth ablation (Section~\ref{sec:abl:contrastive}) asks
whether contrastive supervision on top of cosine alignment helps.

\paragraph{Visual tile budget.}
Block (a) sweeps the tile cap of Section~\ref{sec:method:doc}; the
Fast and HiRes rows are the two deployed variants and the 0-tile
row is a single-448$\times$448-view control.  Removing tiling
costs 5.12 average NDCG@5 relative to HiRes, concentrated on v3.
Going from Fast to HiRes adds 1.76 average points and 3.41 on v3
at $3\times$ the visual tokens, whose deployment cost
Table~\ref{tab:enc} reports.

\paragraph{Training-data scale.}
Block (b) retrains the document encoder on uniform subsamples of
the 1.20\,M-image mixture.  Quality grows monotonically with data
and saturates above 75\,\%; the full mixture is worth about 5
average points over a quarter-scale variant.

\paragraph{Output dimension.}
Because the teacher is trained with Matryoshka representation
learning \cite{li2026qwen3}, any leading prefix of its 4096-d
output is itself a valid embedding, so truncating the target to
768 dimensions is a principled compression rather than a random
projection.  Retraining the document encoder against that
truncated target (block (c)) shrinks the index $5.3\times$
(16.4\,GB $\to$ 3.07\,GB per million documents) but costs 3.19
average NDCG@5 and 4.77 on v3.  We keep 4096-d as the default and
treat 768-d as the option under hard storage constraints.

\paragraph{Query backbone.}
Block (d) retrains the query encoder on BERT-base (110\,M) and
ModernBERT-base (149\,M) under the same protocol.
ModernBERT-base is the strongest at 67.40 average, 1.04 points
above DistilBERT-base, but costs 2.2$\times$ the parameters and
5.1$\times$ the query latency.  Encoding one query at $B=1$
including tokenisation takes 2.07, 4.78 and 10.61\,ms for
DistilBERT-base, BERT-base and ModernBERT-base respectively.
BERT-base trails both on quality at 61.63.  We adopt
DistilBERT-base because 1.04 points is a small price for those
savings.  Index size is unaffected by this choice.

\begin{table}[!t]
  \centering
  \caption{Document- and query-encoder ablations, NDCG@5.  Blocks
    (a) and (b) are end-to-end student$\times$student; block (c) is
    doc-side isolation (student documents against teacher queries);
    block (d) is query-side isolation (student queries against
    teacher documents), whose teacher$\times$teacher ceiling is
    71.05 average.  Absolute values are therefore comparable within
    a block, not across blocks.  $\dagger$ marks the deployed
    default.}
  \label{tab:abl_combined}
  \small
  \setlength{\tabcolsep}{3pt}
  \begin{tabular}{@{}lrrrr@{}}
    \toprule
    Variant & v1 & v2 & v3 & Avg \\
    \midrule
    \multicolumn{5}{@{}l}{\textit{(a) Max tiles (end-to-end)}} \\
    0 (no tiling)                 & 77.57 & 51.29 & 41.00 & 56.62 \\
    2 (Fast)                      & 81.34 & 54.95 & 43.66 & 59.98 \\
    6 (HiRes, default)            & 82.81 & 55.34 & 47.07 & 61.74 \\
    \midrule
    \multicolumn{5}{@{}l}{\textit{(b) Training-data scale (end-to-end)}} \\
    25\,\% (300\,K)               & 78.35 & 49.69 & 40.75 & 56.26 \\
    50\,\% (600\,K)               & 80.97 & 54.44 & 44.58 & 60.00 \\
    75\,\% (900\,K)               & 82.40 & 56.58 & 46.05 & 61.68 \\
    100\,\% (1.20\,M, default)    & 82.81 & 55.34 & 47.07 & 61.74 \\
    \midrule
    \multicolumn{5}{@{}l}{\textit{(c) Doc output dim (doc-side isolation)}} \\
    768-d \;(3.07\,GB/1\,M)       & 81.40 & 58.44 & 45.66 & 61.83 \\
    4096-d \;(16.4\,GB/1\,M)$^\dagger$
                                  & 83.72 & 60.92 & 50.43 & 65.02 \\
    \midrule
    \multicolumn{5}{@{}l}{\textit{(d) Query backbone (query-side isolation)}} \\
    DistilBERT-base (70\,M)$^\dagger$
                                  & 84.68 & 64.30 & 50.09 & 66.36 \\
    BERT-base (110\,M)            & 81.81 & 58.52 & 44.55 & 61.63 \\
    ModernBERT-base (149\,M)      & 85.36 & 65.40 & 51.43 & 67.40 \\
    \bottomrule
  \end{tabular}
\end{table}

\subsection{Contrastive supervision}
\label{sec:abl:contrastive}
Our recipe trains the two students independently, with no
cross-modal interaction in the loss.  This ablation adds a joint
refinement step that couples them through a contrastive term over
paired $(q,d)$ mini-batches.  Starting from a cosine-distilled
checkpoint of the same 524\,M architecture at a 768-d projection
head (the 768-d counterpart of the 4096-d encoder, with the same
encoders, data, tiling, and teacher), we train both students
jointly for one further epoch on the same mixture under
\begin{equation}
  \mathcal{L} = \alpha\,\mathcal{L}_d + \beta\,\mathcal{L}_q
              + \gamma\,\mathcal{L}_{\mathrm{rank}},
  \label{eq:ablloss}
\end{equation}
where $\mathcal{L}_d$ and $\mathcal{L}_q$ are the cosine alignment
losses of Eqs.~\ref{eq:ld}--\ref{eq:lq} and $\alpha=\beta=1.0$.
For a mini-batch $\mathcal{B}$ of paired query-document examples,
let
$s_{qd}=\langle f_q(q),\, f_d(d)\rangle$ denote the student
similarity and
$\tilde{s}_{qd}=\langle T(q),\, T(d)\rangle$ the teacher
similarity between $q$ and $d$.  We consider two choices for
$\mathcal{L}_{\mathrm{rank}}$:
\begin{align}
  \mathcal{L}_{\mathrm{InfoNCE}}
    &= -\!\!\sum_{(q,d)\in\mathcal{B}}\!\!
       \log
       \frac{\exp(s_{qd}/\tau)}
            {\sum_{d'\in\mathcal{B}}\exp(s_{qd'}/\tau)},
       \label{eq:infonce}\\
  \mathcal{L}_{\mathrm{KL}}
    &= \sum_{q\in\mathcal{B}}
       \mathrm{KL}\!\big(\,\sigma_\tau(\tilde{s}_q)\,\|\,
                          \sigma_\tau(s_q)\,\big),
       \label{eq:klrank}
\end{align}
with temperature $\tau=0.02$ and $\sigma_\tau$ the row-wise softmax
of similarities to all documents in $\mathcal{B}$.  The refinement
uses AdamW, effective batch 256, peak learning rate $2\times
10^{-5}$, warm-up fraction 0.05, and the dynamic-tiling rule of
DistilVDR-HiRes on two H200 GPUs.  Absolute NDCG@5 in
Table~\ref{tab:abl_contrastive} is below
Section~\ref{sec:exp:quality} because the projection head outputs
768 dimensions instead of 4096; the relative effect of $\gamma$
is unaffected by this offset.

Within this 1-epoch joint refinement, adding contrastive
supervision yields no NDCG@5 improvement at any $\gamma$ we
tried.  InfoNCE shows a mild downward drift with increasing
$\gamma$, from $-0.12$ at $\gamma=0.5$ to $-0.50$ at
$\gamma=2.0$.  KL is flat across all weights
($\pm 0.14$).  The asymmetry is consistent with the two losses'
geometric content: InfoNCE pushes the student toward a uniform
separator over in-batch negatives, orthogonal to the
teacher-calibrated geometry already enforced by
$\mathcal{L}_d+\mathcal{L}_q$; KL anchors to the teacher
distribution and is approximately redundant with cosine alignment.
The refinement budget is roughly one third of the document
encoder's training budget (Table~\ref{tab:recipe}), enough
that any positive contribution from the contrastive term
should register a positive drift; what we observe is flat
or mildly negative.  The direction is consistent with the
from-scratch loss ablation of NanoVDR~\cite{liu2026nanovdr},
which reports pointwise cosine alignment strictly outperforming
both KL-ranking and hard-label InfoNCE distillation under the
same teacher family.

Every configuration here starts from the same distilled
checkpoint, so the experiment covers the refinement stage only and
does not compare distillation and contrastive learning as
alternative ways of training the same architecture.  Within that
scope it supports the pure cosine objective of
Section~\ref{sec:method:loss} as a strong default.

\begin{table}[!t]
  \centering
  \caption{Contrastive-supervision ablation under
    Eq.~\ref{eq:ablloss}.  All rows are one-epoch joint refinements
    from the cosine-distilled 768-d sibling checkpoint; absolute
    NDCG@5 is below Section~\ref{sec:exp:quality} because of the
    768-d output.}
  \label{tab:abl_contrastive}
  \small
  \begin{tabular}{@{}lrrrr@{}}
    \toprule
    Refinement objective                                       & v1    & v2    & v3    & Avg   \\
    \midrule
    $\gamma = 0$ (cosine only)                                 & 79.90 & 53.23 & 42.22 & 58.45 \\
    \midrule
    \;\;$+\, 0.5\,\mathcal{L}_{\mathrm{InfoNCE}}$              & 80.47 & 52.08 & 42.44 & 58.33 \\
    \;\;$+\, 1.0\,\mathcal{L}_{\mathrm{InfoNCE}}$              & 80.48 & 51.80 & 42.41 & 58.23 \\
    \;\;$+\, 2.0\,\mathcal{L}_{\mathrm{InfoNCE}}$              & 80.25 & 51.31 & 42.30 & 57.95 \\
    \midrule
    \;\;$+\, 0.5\,\mathcal{L}_{\mathrm{KL}}$                   & 80.32 & 52.31 & 42.30 & 58.31 \\
    \;\;$+\, 1.0\,\mathcal{L}_{\mathrm{KL}}$                   & 80.20 & 52.55 & 42.34 & 58.36 \\
    \;\;$+\, 2.0\,\mathcal{L}_{\mathrm{KL}}$                   & 80.31 & 52.54 & 42.44 & 58.43 \\
    \bottomrule
  \end{tabular}
\end{table}

\section{Conclusion}
\label{sec:conc}

We presented DistilVDR, a 524\,M end-to-end visual document
retriever obtained by independent cosine distillation of both
encoders from a single 8\,B vision-language teacher under an
asymmetric encoder-only design.  Its two deployment variants share
encoders and training and differ only in the document-side
visual-tile budget: HiRes retains 86.9\,\% of the teacher and
leads every reproduced sub-1\,B baseline on ViDoRe v3, Fast
retains 84.4\,\% at a $3\times$ smaller visual-token budget, and
both index an order of magnitude faster at a 15.6 times smaller
footprint.
Measured end to end under one evaluation and one profiling
pipeline, distilling both towers of a strong single-vector
vision-language teacher reaches a better quality-and-cost
operating point at the 500\,M scale than any released small
multi-vector retriever we could reproduce.

\section*{Limitations}

\paragraph{From an experimental perspective.}
Every comparison we report is between deployable systems rather
than between training recipes.
Every baseline in Table~\ref{tab:main} is an official public
release evaluated as released; none is retrained under a budget
matched to ours, and our teacher, our 1.20\,M-image mixture, and
our architecture all differ from theirs, so the margin cannot be
attributed to any single one of them.  The control that would
isolate distillation itself, the same 524\,M architecture trained
contrastively from scratch on the same mixture with hard-negative
mining and a comparable compute budget, is the principal
experiment we did not run; the refinement experiment of
Section~\ref{sec:abl:contrastive} starts from an already-distilled
checkpoint and does not substitute for it.  We likewise use
exactly one teacher and do not sweep teacher scale or family, so
we cannot say how much quality a weaker teacher would cost or how
much a stronger one would add; because targets are
cached once and student training never runs the teacher, that
sweep is a cheap follow-up.  Both variants also stay 7--10 points behind the
strongest 4--8\,B multi-vector references of
Table~\ref{tab:main}, with the largest gap on
ViDoRe v3 (47.07 and 43.66 against 57.57--59.00), and because our
experiments vary the scoring mechanism only together with model
scale, they cannot separate late interaction from capacity at the
top of that table.  Finally, every retrieval number comes from the
22 ViDoRe datasets.  ViDoRe spans three difficulty levels, six
languages, and eight professional domains, but not enterprise
corpora with in-house layouts, scanned or OCR-degraded pages, or
production query distributions; our deployment costs are measured
directly and transfer, our quality numbers may not.

\paragraph{From a methodological perspective.}
The student is bounded by the space it is trained into.  Both
towers only reproduce a frozen teacher's embeddings, so any
systematic weakness of Qwen3-VL-Embedding-8B propagates to
DistilVDR, and nothing in the objective lets a student exceed its
teacher on the side it replaces; Table~\ref{tab:gap_decomp} is
consistent with this, as no student-side substitution improves on
the teacher$\times$teacher oracle.  Distilling from a richer
signal, such as a cross-encoder relevance score or a multi-vector
teacher, is what would raise that ceiling.  Three further choices
are deliberate simplifications.  The two students are trained
independently against cached targets, which parallelises cleanly
but leaves unused the cross-modal coupling between query and
document manifolds that the teacher implicitly carries; a joint
scheme with explicit query-document interaction during
distillation is the natural extension.  The tiling rule adapts the
grid to page aspect ratio but not to page content, and $T_{\max}$
is fixed at release time, so a dense small-text page and a
single-figure page receive the same visual-token budget; our two
variants sample two points on that trade-off rather than resolve
it, and a content-adaptive budget predicted from a cheap
page-complexity estimate would spend tokens where they are
needed.  And the query tower is text-only,
trained on English plus five Latin-script European languages
obtained through a MarianMT translation pipeline, so
image-conditioned queries and non-Latin scripts such as Chinese,
Japanese, and Arabic are out of scope.  One deployment-side choice
is also unexplored.  We store one uncompressed 4096-d
\texttt{float32} vector per document.
Table~\ref{tab:abl_combined} measures a single compression axis,
the 768-d Matryoshka prefix (3.07\,GB per million documents,
$-3.19$ NDCG@5 average); scalar quantisation, product
quantisation, and binary hashing are orthogonal to our
contribution and untested here, so the reported footprint should
be read as an upper bound.

\bibliography{references}

\appendix

\section{ViDoRe Benchmark Details}
\label{app:vidore}

The Visual Document Retrieval (ViDoRe) benchmark comprises three
progressively challenging versions
\cite{faysse2025colpali,mace2025vidore,loison2026vidore}.  We
evaluate on the full 22-dataset suite.  Table~\ref{tab:vidore-datasets}
lists each dataset together with its document corpus language, query
languages, and the source of its queries.

\paragraph{v1 \cite{faysse2025colpali}.}
Released with ColPali, v1 contains 10 English- and French-language
datasets.  Five are sourced from established visual QA benchmarks
(DocVQA, ArXivQA, InfoVQA, TatDQA, TabFQuAD) with human-authored
queries; five use queries generated by Claude-3 Sonnet over curated
document collections.  State-of-the-art models now exceed 90 NDCG@5
on v1, indicating saturation.

\paragraph{v2 \cite{mace2025vidore}.}
Designed to address v1's saturation, v2 introduces 4 datasets with
queries in up to four languages (English, French, Spanish, German).
Queries are generated through a blind contextual process: annotators
receive only document metadata, which reduces extractive bias and
requires cross-document reasoning.  The fourth dataset (ESG Reports
Human-Labeled) provides expert human annotations over the same ESG
corpus.

\paragraph{v3 \cite{loison2026vidore}.}
The most comprehensive version, v3 provides 8 public datasets across
six languages (the v2 languages plus Italian and Portuguese).
Annotation combined LLM-synthesised queries with expert review.  Each
query carries page-level relevance rankings, bounding-box
annotations, and multilingual translations.  The eight domains
span enterprise scenarios from finance to physics, with document
corpora in English or French.

\begin{table*}[h]
  \centering
  \small
  \setlength{\tabcolsep}{4pt}
  \caption{All 22 ViDoRe evaluation datasets.  ``Doc'' is the
    document corpus language.  ``Query'' is the languages in which
    queries are released.  ``Source'' indicates how the queries were
    obtained: H = human-authored, L = LLM-generated, L+H =
    LLM-generated with expert review.  The six languages of v3 are
    English, French, Spanish, German, Italian, and Portuguese.}
  \label{tab:vidore-datasets}
  \begin{tabular}{@{}lcllcl@{}}
    \toprule
    Dataset                & Ver. & Document domain          & Doc & Query           & Source \\
    \midrule
    DocVQA                 & v1   & Industrial documents     & EN  & EN              & H     \\
    ArXivQA                & v1   & Scientific papers        & EN  & EN              & H     \\
    InfoVQA                & v1   & Infographics             & EN  & EN              & H     \\
    TatDQA                 & v1   & Financial tables         & EN  & EN              & H     \\
    TabFQuAD               & v1   & Tables in French PDFs    & FR  & FR              & H     \\
    SyntheticDocQA-AI      & v1   & AI documents             & EN  & EN              & L     \\
    SyntheticDocQA-Energy  & v1   & Energy sector reports    & EN  & EN              & L     \\
    SyntheticDocQA-Gov.    & v1   & Government reports       & EN  & EN              & L     \\
    SyntheticDocQA-Hlt.    & v1   & Healthcare documents     & EN  & EN              & L     \\
    ShiftProject           & v1   & Environmental reports    & FR  & FR              & L     \\
    \midrule
    ESG Reports            & v2   & ESG / sustainability     & EN  & EN/FR/ES/DE     & L+H   \\
    Biomedical Lectures    & v2   & Biomedical slides        & EN  & EN/FR/ES/DE     & L+H   \\
    Economics Reports      & v2   & Economics reports        & EN  & EN/FR/ES/DE     & L+H   \\
    ESG Reports (Human)    & v2   & ESG / sustainability     & EN  & EN              & H     \\
    \midrule
    Finance-EN             & v3   & US annual reports        & EN  & 6 languages     & L+H   \\
    Finance-FR             & v3   & French annual reports    & FR  & 6 languages     & L+H   \\
    Computer Science       & v3   & CS textbooks             & EN  & 6 languages     & L+H   \\
    HR                     & v3   & EU HR reports            & EN  & 6 languages     & L+H   \\
    Energy                 & v3   & French energy reports    & FR  & 6 languages     & L+H   \\
    Industrial             & v3   & USAF technical orders    & EN  & 6 languages     & L+H   \\
    Pharmaceutical         & v3   & FDA reports              & EN  & 6 languages     & L+H   \\
    Physics                & v3   & French physics lectures  & FR  & 6 languages     & L+H   \\
    \bottomrule
  \end{tabular}
\end{table*}

\section{Training Data Composition and Preprocessing}
\label{app:data}

The document encoder training mixture combines three groups of
publicly released datasets.  Table~\ref{tab:data-detail} gives the
full per-source breakdown, including the query-encoder mixture.

\paragraph{Base sources.}
The first group reproduces the 711\,K base mixture released by
NanoVDR \cite{liu2026nanovdr}: VisRAG-Synthetic and VisRAG-InDomain
\cite{yu2025visrag}, the ColPali training set
\cite{faysse2025colpali}, and VDR-Multilingual (English, French,
German, Spanish, Italian).  The query encoder uses the same 711\,K
queries together with the 778\,K Helsinki-NLP MarianMT translations
released with NanoVDR, for a total of 1.49\,M paired queries.

\paragraph{Multi-domain supplement.}
The second group is the union of fourteen domain-specific document
collections released by Racineai.  These collections are aggregated
under the \texttt{racineai/VDR\_MEGA\_2} release and individually
contributed under permissive licences.  We deduplicate against the
711\,K base by perceptual hash and against the ViDoRe v1/v2/v3
evaluation corpora.  The resulting 454\,K-image subset is used on
the document side only.

\paragraph{Finance supplement.}
The third group adds 31.7\,K finance-domain document images from
\texttt{sujet-ai/Sujet\allowbreak-Finance\allowbreak-Vision\allowbreak-10k} and
\texttt{DocReRank/\allowbreak FinHNQue\allowbreak-Finance\allowbreak Hard\allowbreak Negative\allowbreak Queries}, again
deduplicated by perceptual hash against base and evaluation
corpora.

\paragraph{Preprocessing pipeline.}
All training images are loaded as PIL RGB and processed by the
encoder pipeline of Section~\ref{sec:method:doc} (dynamic tiling,
patch encoding, projection).  Deduplication is performed offline
with the \texttt{imagehash} library at perceptual-hash distance 0,
which removes only exact visual duplicates.  Teacher embeddings are
precomputed once over the deduplicated mixture and cached as
\texttt{float32} arrays on disk, so the teacher does not run during
student training.

\begin{table}[h]
  \centering
  \small
  \setlength{\tabcolsep}{4pt}
  \caption{Training data composition.  HuggingFace identifiers are
    hyperlinked; the organisation prefix is omitted in the second
    and third groups.}
  \label{tab:data-detail}
  \begin{tabular}{@{}p{5.1cm}r@{}}
    \toprule
    Source                                                & Samples \\
    \midrule
    \multicolumn{2}{@{}l}{\textit{Base mixture (711\,K)}}            \\
    \href{https://huggingface.co/datasets/openbmb/VisRAG-Ret-Train-Synthetic-data}{%
      \texttt{VisRAG-Ret-Train-Synthetic}}~\cite{yu2025visrag}       & 234\,K  \\
    \href{https://huggingface.co/datasets/openbmb/VisRAG-Ret-Train-In-domain-data}{%
      \texttt{VisRAG-Ret-Train-In-domain}}~\cite{yu2025visrag}       &  94\,K  \\
    \href{https://huggingface.co/datasets/vidore/colpali_train_set}{%
      \texttt{colpali\_train\_set}}~\cite{faysse2025colpali}         & 109\,K  \\
    \href{https://huggingface.co/datasets/llamaindex/vdr-multilingual-train}{%
      \texttt{vdr-multilingual-train}}                              & 275\,K  \\
    \midrule
    \multicolumn{2}{@{}l}{\textit{Multi-domain supplement (Racineai, deduplicated)}}  \\
    \href{https://huggingface.co/datasets/racineai/VDR_MEGA_2}{%
      \texttt{racineai/VDR\_*}} (14 sub-sources)                     & 454\,K \\
    \midrule
    \multicolumn{2}{@{}l}{\textit{Finance supplement (31.7\,K)}}    \\
    \href{https://huggingface.co/datasets/sujet-ai/Sujet-Finance-Vision-10k}{%
      \texttt{Sujet-Finance-Vision-10k}}                            & 9.8\,K  \\
    \href{https://huggingface.co/datasets/DocReRank/FinHNQue-FinanceHardNegativeQueries}{%
      \texttt{FinHNQue}}                                            & 21.9\,K \\
    \midrule
    \textbf{Document encoder total}                       & \textbf{1.20\,M} \\
    \midrule
    \multicolumn{2}{@{}l}{\textit{Query encoder mixture}}            \\
    NanoVDR query training set~\cite{liu2026nanovdr}      & 1.49\,M \\
    \bottomrule
  \end{tabular}
\end{table}

\section{Baselines}
\label{app:baselines}

At the sub-1\,B scale the multi-vector retrievers are colSmol-256M
and colSmol-500M \cite{vidore2024colsmol},
SauerkrautLM-ColLFM2-450M \cite{vago2025collfm2}, and
ColModernVBERT \cite{teiletche2025modernvbert}; the single-vector
retrievers are SigLIP2-L
\cite{tschannen2025siglip2multilingualvisionlanguage} and
BiModernVBERT \cite{teiletche2025modernvbert}.  At the 2--3\,B
scale we include the single-vector DSE-Qwen2 \cite{ma2024unifying}
and Qwen3-VL-Embedding-2B \cite{li2026qwen3}, and the multi-vector
ColPali v1.3 \cite{faysse2025colpali}.  At the 4--8\,B scale we
include Tomoro-ColQwen3-4B and Tomoro-ColQwen3-8B
\cite{tomoro2025colqwen3} and ColNomic-7B \cite{nomic2025colnomic}.
Qwen3-VL-Embedding-8B \cite{li2026qwen3} is our distillation
teacher and is also the single-vector oracle.

\section{Training Recipe}
\label{app:recipe}

Table~\ref{tab:recipe} lists the hyperparameters and hardware for
both students under the cosine alignment objective of
Eqs.~\ref{eq:ld}--\ref{eq:lq}.  The two document-encoder variants
(Fast and HiRes) share this recipe exactly and differ only in
$T_{\max}$.  Per-epoch loss curves for all three students are
reported as a convergence diagnostic in Appendix~\ref{app:loss}.

\begin{table}[h]
  \centering
  \caption{Training recipe under the cosine alignment objective
    (Eq.~\ref{eq:ld}--\ref{eq:lq}).}
  \label{tab:recipe}
  \small
  \setlength{\tabcolsep}{4pt}
  \begin{tabular}{@{}lll@{}}
    \toprule
    Hyperparameter & Doc enc. & Query enc. \\
    \midrule
    Trainable parameters & 454\,M & 70\,M \\
    Optimizer & AdamW & AdamW \\
    Peak LR & $1\!\times\!10^{-3}$ & $5\!\times\!10^{-4}$ \\
    LR schedule & \multicolumn{2}{l}{one-cycle, 3\% warmup} \\
    Effective batch & 256 & 512 \\
    Epochs & 3 & 15 \\
    Hardware & 2$\times$ H200 & 2$\times$ H200 \\
    \bottomrule
  \end{tabular}
\end{table}

\section{Dynamic Tiling Algorithm}
\label{app:tiling}

Algorithm~\ref{alg:tile} formalises the InternVL-V2 dynamic tiling
rule \cite{chen2025expandingperformanceboundariesopensource} used by
the document encoder of Section~\ref{sec:method:doc}.  The rule
picks an aspect-ratio-matched grid layout from the candidate set
$\mathcal{G} = \{(p, q) : p, q \in \mathbb{Z}_{\geq 1},\; n_{\min}
\leq p \cdot q \leq n_{\max}\}$, resizes the image to fit that grid
at the encoder's native tile resolution, splits the resized image
into non-overlapping tiles, and optionally appends a single
thumbnail at the same resolution to preserve global layout context.
Tie-breaking among candidates with equal aspect distance prefers
the higher-resolution grid when the original image area is large.
We use $n_{\min} = 1$, $n_{\max} = 6$, tile size $s = 448$, and
enable the thumbnail.  The maximum sequence length per document is
$(n_{\max} + 1) \cdot s^2 / p^2 = 7 \cdot 1024 = 7168$ patch tokens
at the InternViT-300M patch size of $p = 14$, which fits inside
ModernBERT-base's 8192-token context window without truncation.

\begin{algorithm}[h]
  \caption{Dynamic Tiling}
  \label{alg:tile}
  \begin{algorithmic}[1]
    \Require image $I$ of size $(W, H)$; $n_{\min}$, $n_{\max}$;
             tile size $s$; thumbnail flag $t$
    \Ensure list of tiles, each $s \times s$
    \State $r \gets W / H$ \Comment{input aspect ratio}
    \State $\mathcal{G} \gets \{ (p, q) : p, q \in \mathbb{Z}_{\geq 1},\;
             n_{\min} \leq p\,q \leq n_{\max} \}$
    \State $(p^*, q^*) \gets \arg\min_{(p,q) \in \mathcal{G}}
             | r - p / q |$ \Comment{ties broken toward higher resolution}
    \State $I' \gets \text{Resize}(I,\; (p^* s,\; q^* s))$
    \State $\mathcal{T} \gets$ split $I'$ into $p^* q^*$
             non-overlapping $s \times s$ tiles, left-to-right then top-to-bottom
    \If{$t$ and $p^* q^* > 1$}
      \State $\mathcal{T} \gets \mathcal{T} \cup \{\text{Resize}(I,\;(s, s))\}$
        \Comment{thumbnail for global context}
    \EndIf
    \State \Return $\mathcal{T}$
  \end{algorithmic}
\end{algorithm}

\section{Training Loss Curves}
\label{app:loss}

\begin{figure}[!t]
  \centering
  \includegraphics[width=0.85\columnwidth]{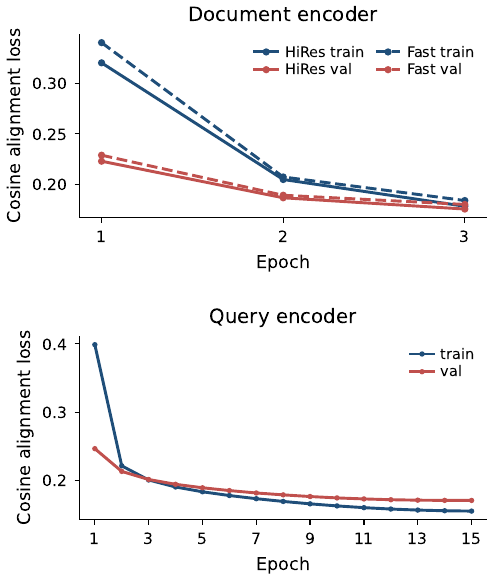}
  \caption{Per-epoch train (blue) and validation (red) loss for
    the document students DistilVDR-HiRes (solid) and
    DistilVDR-Fast (dashed) on top, and the shared query student
    on bottom.}
  \label{fig:loss}
\end{figure}

Figure~\ref{fig:loss} reports per-epoch train and validation
cosine-alignment loss for the two document-encoder variants
(DistilVDR-HiRes and DistilVDR-Fast) and for the shared query
encoder under the recipe of Table~\ref{tab:recipe}.  All curves
decrease smoothly to a low plateau with a small
train/validation gap, indicating that every student reaches a
steady-state fit to its teacher target within the allotted
epoch budget.

\section{Efficiency Benchmark Protocol}
\label{app:effbench}

\paragraph{Hardware and software stack.}
A single NVIDIA H200 GPU (141\,GB HBM3e) on an Intel Xeon Platinum
8480+ node with 32 CPU cores and 128\,GB RAM.  Software: PyTorch
2.5, CUDA 12.6, transformers $\geq$\,4.46.  All encoders run in
bfloat16.

\paragraph{Attention backend per encoder.}
Every encoder runs under a flash-attention-family
kernel~\cite{dao2022flashattention,dao2024flashattention}.  For all
baselines and DistilVDR's ModernBERT text backbone this kernel is
reached through PyTorch's sdpa backend, which dispatches to
FlashAttention-2 in transformers $\geq$\,4.46; calling
\texttt{attn\_implementation="flash\_attention\_2"} explicitly
changes throughput by under $\pm 1$\,docs/sec.  DistilVDR's
InternViT visual encoder does not register an sdpa implementation
with transformers, but its remote modeling code ships its own
native flash-attention path, enabled by default via the
\texttt{use\_flash\_attn=True} config flag; this is the
configuration reported in Table~\ref{tab:enc}.

\paragraph{Per-baseline deviations.}
Two baselines deviate from the default bf16$+$sdpa setup.
BiModernVBERT is loaded in float32 because the canonical
evaluation pipeline observes that bf16 mean-pooling over the
1\,000+ visual tokens emitted by its ModernBERT backbone incurs a
several-NDCG drop; we follow the released configuration and use
float32 on both the document and query sides.  DSE-Qwen2 follows its
released chat-template inference path, wrapping each document
image in a \texttt{user} message with a fixed
$680 \times 680$ resized resolution and reading the last hidden
state at the trailing \texttt{<|endoftext|>} token as the
document embedding; we use the model's released
\texttt{prepare\_inputs\_for\_generation} entry point with
\texttt{use\_cache=False} to obtain that hidden state in a single
forward pass.

\paragraph{Conservative attention-backend reference.}
For transparency we also benchmark a conservative DistilVDR variant
that forces InternViT to eager attention
(\texttt{use\_flash\_attn=False}).  At $B=8$, DistilVDR-Fast drops
to 31.91 docs/sec (VRAM 5.73\,GB) and DistilVDR-HiRes to 9.27
docs/sec (11.56\,GB); even under this configuration DistilVDR-Fast
remains faster than every sub-1\,B multi-vector baseline.

\paragraph{Measurement protocol.}
Query latency is the mean over 20 queries at $B=1$ after 3 warmup
queries, with \texttt{cuda.synchronize()} bracketing the forward
call and tokenisation included.  Document throughput and peak
VRAM are measured at fixed batch $B=8$ on a 100-page synthetic
corpus after 3 warmup forward passes; multi-vector baselines run
their official per-token pooling and projection so that the
per-document tensor written to the index is fully formed.  Index
size is computed analytically from each baseline's released
storage format ($T$ tokens per document at dimension $d$ in
float16 for multi-vector retrievers; one $d$-dimensional float32
vector per document for single-vector retrievers).  CPU scoring
latency is the mean over 20 runs on a single thread
($\texttt{torch.set\_num\_threads(1)}$) against a synthetic
10\,000-candidate corpus in the appropriate storage format; the
single-vector path is a dot product, the multi-vector path is the
MaxSim routine over 32 query tokens.

\end{document}